\documentclass[a4paper,12pt]{article}
\usepackage{epsfig}
\usepackage{amssymb}
\usepackage{amsfonts}

\newskip\humongous \humongous=0pt plus 1000pt minus 1000pt

\newif\ifdtup

\catcode`\@=11

\@addtoreset{equation}{section}
\def\theequation{\thesection.\arabic{equation}}

\def\@normalsize{\@setsize\normalsize{15pt}\xiipt\@xiipt
\abovedisplayskip 14pt plus3pt minus3pt%
\belowdisplayskip \abovedisplayskip
\abovedisplayshortskip \z@ plus3pt%
\belowdisplayshortskip 7pt plus3.5pt minus0pt}

\def\small{\@setsize\small{13.6pt}\xipt\@xipt
\abovedisplayskip 13pt plus3pt minus3pt%
\belowdisplayskip \abovedisplayskip
\abovedisplayshortskip \z@ plus3pt%
\belowdisplayshortskip 7pt plus3.5pt minus0pt
\def\@listi{\parsep 4.5pt plus 2pt minus 1pt
     \itemsep \parsep
     \topsep 9pt plus 3pt minus 3pt}}

\relax

\catcode`@=12

\catcode`\@=11

\def\section{\@startsection{section}{1}{\z@}{3.5ex plus 1ex minus
   .2ex}{2.3ex plus .2ex}{\large\bf}}

\def\thesection{\arabic{section}}
\def\thesubsection{\arabic{section}.\arabic{subsection}}

\def\appendix{\setcounter{section}{0}
 \def\thesection{Appendix \Alph{section}}
 \def\thesubsection{\Alph{section}.\arabic{subsection}}
 \def\theequation{\Alph{section}.\arabic{equation}}}

\begin{document}

%%%%%%Define some new commands and  macros
\newcommand{\beq}{\begin{equation}}
\newcommand{\eeq}{\end{equation}}
\newcommand{\bea}{\begin{eqnarray}}
\newcommand{\eea}{\end{eqnarray}}
\newcommand{\beas}{\begin{eqnarray*}}
\newcommand{\eeas}{\end{eqnarray*}}
\newcommand{\defi}{\stackrel{\rm def}{=}}
\newcommand{\non}{\nonumber}
\newcommand{\bquo}{\begin{quote}}
\newcommand{\enqu}{\end{quote}}
\newcommand{\mat}{\mathbf}
%%%%%%%%%%%%%%%%%%%%%%%%%%%%%%%%%% definitions
\def\de{\partial}
\def\Tr{ \hbox{\rm Tr}}
\def\const{\hbox {\rm const.}}
\def\o{\over}
\def\im{\hbox{\rm Im}}
\def\re{\hbox{\rm Re}}
\def\bra{\langle}\def\ket{\rangle}
\def\Arg{\hbox {\rm Arg}}
\def\Re{\hbox {\rm Re}}
\def\Im{\hbox {\rm Im}}
\def\diag{\hbox{\rm diag}}
\def\longvert{{\rule[-2mm]{0.1mm}{7mm}}\,}
\def\Z{\mathbb Z}
\def\N{{\cal N}}
\def\tq{{\widetilde q}}
\def\W{{\cal W}}
\def\tQ{{\widetilde Q}}
\def\dag{{}^{\dagger}}
\def\p{{}^{\,\prime}}
\def\a{\alpha}
\def\Tr{ \hbox{\rm Tr}}
\def\tM{{\widetilde M}}
\def\tm{{\widetilde m}}
\def\T{{\cal T}}
\def\t{T}
\def\J{{\cal J}}

\begin{titlepage}
\begin{flushright}
hep-th/0411075\\
\end{flushright}

\bigskip

\begin{center}
{\Large

{\bf  
The Holomorphic Tension of Vortices
}

 } 
\end{center}

\renewcommand{\thefootnote}{\fnsymbol{footnote}}
\bigskip
\begin{center}
{\large   Stefano Bolognesi }
 \vskip 0.20cm
\end{center}

\begin{center}
{\it      \footnotesize
Scuola Normale Superiore - Pisa, Piazza dei Cavalieri 7, Pisa, Italy \\
\vskip 0.10cm
and\\
\vskip 0.10cm
Istituto Nazionale di Fisica Nucleare -- Sezione di Pisa, \\
Via Buonarroti 2, Ed. C, 56127 Pisa,  Italy   \\  }
\vskip 0.15cm
s.bolognesi@sns.it\\
\end {center}

\setcounter{footnote}{0}

\bigskip
\bigskip

\noindent  
\begin{center} {\bf Abstract} \end{center}
We study the tension of vortices in ${\cal N}=2$ SQCD broken to ${\cal N}=1$ by a superpotential $W(\Phi)$, in color-flavor locked vacua. The tension can be written as  $T=T_{BPS} + T_{non\, BPS}$. The BPS tension is equal to $4\pi|\T|$  where we call $\T$  the {\it holomorphic tension}. This is directly related to the central charge of the supersymmetry algebra. Using the tools of the Cachazo-Douglas-Seiberg-Witten solution we compute the holomorphic tension as a holomorphic function of the couplings, the mass and the dynamical scale: $\T=\sqrt{{W\p}^2+f}$. A first approximation is given using the generalized Konishi anomaly in the semiclassical limit.  The full quantum corrections are computed in the strong coupling regime using the factorization equations that relate the $\N=2$ curve to the $\N=1$ curve.   Finally we study the limit in which the non-BPS contribution  can be neglected because small with respect to the BPS one.  In the case of linear superpotential the non-BPS contribution vanishes exactly and the holomorphic tension gets no quantum corrections.

\vfill

\begin{flushleft}
November, 2004
\end{flushleft}

\end{titlepage}

\bigskip

\hfill{}

\section{Introduction}

In the last years important progress has been made in the study of strongly coupled dynamics of various supersymmetric gauge theories. In particular, the understanding of supersymmetric constraints, the study of solitonic objects and various kind of dualities, have led to many exact results.

The works of Seiberg and Witten \cite{SW1,SW2} were the starting point for the understanding of the exact low energy dynamics of $\N=2$ theories \cite{curvesT}. In particular they found an  exact formula for the  mass of the BPS saturated dyons:
\beq
\label{BPSmass}
M_{BPS}=\sqrt{2} |a n_e +a_D n_m|\ ,
\eeq 
where $n_e$ and $n_m$ are respectively the electric and the magnetic charge.

In this paper we find something similar for vortices (the Abrikosov-Nielsen-Olesen flux tubes \cite{NO}) in $\N=2$ SQCD broken to $\N=1$ by a superpotential $W(\Phi)$. When there is a color-flavor locking some flavors become massless in the $\N=2$ theory.  Due to the presence of the superpotential, these flavors condense and create a vortex solution. The tension of the vortex can be written as a BPS tension plus a non-BPS contribution
\beq
\label{claim}
T=T_{BPS} + T_{non\, BPS}\ ,
\eeq
where 
\beq
T_{BPS}=4\pi|\T| 
\eeq 
 and we call $\T$ the {\it  holomorphic tension}.    In a recent work \cite{RioandJarah} these vortices were studied in the semiclassical limit ($m>>\Lambda$), where the tension is the classical one plus quantum corrections  that depend on the dynamical scale $\Lambda$
\beq
\T=W\p(m)+{\cal O}\left(\frac{\Lambda}{m}\right)\ .
\eeq
In the present paper we compute the quantum corrections to the holomorphic tension and our result, in the semiclassical limit, is  a resummation of infinite instanton contributions. We are not able to compute the non-BPS contribution $T_{non \,BPS}$.

The formula (\ref{BPSmass}) has a deep relation with the central charge of the $\N=2$ superalgebra. When the mass saturates the BPS bound, half of the supersymmetries are unbroken and this prevents quantum corrections. Something similar happens with vortices in an $\N=1$ theory.   As studied in \cite{GS}, if Lorentz invariance is broken by a vortex configuration, one can introduce a central charge in the $\N=1$ superalgebra. This central charge is essentialy the holomorphic tension $\T$.

Recently the work of Dijkgraaf and Vafa \cite{DV} has  opened the way to new discoveries about the $\N=1$ non perturbative dynamics. In particular in \cite{CDSW,CSW2,CSW} the $\N=2$ SQCD broken to $\N=1$ by a superpotential has been studied. The tools developed in these works will be essential in this paper.

Thanks to a generalized version of the Konishi anomaly one can compute all the expectation values of the operators in the chiral ring of the theory. The generators of the chiral ring are the power expansion in $z$ of some quantities that are usually denoted  $T(z)$, $R(z)$ and $M(z)$. These are differential forms on the Riemann surface $\Sigma_{\N=1}$ defined by the equation
\beq
\label{curva1intr}
 {y_{m}}^2={W\p}^2(z)+f(z)\ ,
\eeq 
where $f(z)$ is a polynomial that depends on the vacuum.
In the semiclassical limit, where $m>> \Lambda$, the low energy degrees of freedom are the color locked quarks and  so we expect the tension to be given by their condensate
\beq
\label{conjecturefirst}
\T=- \tQ  Q \ .
\eeq
This condensate belongs to the chiral ring and can be computed exactly by the residue
\beq
\label{aaaaaaaaaaaaaaa}
 \tQ Q=\frac{1}{2\pi i} \oint_{\infty} M(z) dz = -\frac{1}{2} \left.(W\p + \sqrt{{W\p}^2+f}) \right|_{z=m} \ ,
\eeq
where $m$ is the bare mass of the locked flavor.

Even if the generalized anomaly gives the condensate $Q\tQ$ for every value of $m$, only in the semiclassical region these are the low energy degrees of freedom that create the vortex. Thus we expect that  (\ref{conjecturefirst}) will get other corrections and to get them we perform a computation in the strong coupling regime. The strong coupling is a regime in which the superpotential can be considered a small perturbation to the $\N=2$ theory. The following factorization equations give the relation between the Seiberg-Witten curve $\Sigma_{\N=2}$ and the  curve $\Sigma_{\N=1}$: 
\beq
\label{factoroneintr}
{y}^2=P_{N_c}(z)^2-\Lambda^{2N_c-N_f}\prod_{I=1}^{N_f}(z-m_I)=
F_{2n}(z)H_{N_c-n}(z)^2 
\eeq
and
\beq
\label{factortwointr}
{y_{m}}^2=W\p(z)^2+f(z)={g_{k}}^2 F_{2n}(z) Q_{k-n}(z)^2 \ .
\eeq
This factorization has the following interpretation. The $\N=2$ low energy has $N_c$ $U(1)$ factors and $N_c-n$ of them are coupled to massless particles (this is seen in (\ref{factoroneintr}) by the $N_c-n$ double roots).  One of these double roots is $\tm$, that in the semiclassical limit becomes the bare mass of the locked flavor $m$.  The computation of the tension gives
\beq
\label{resultintro}
\T=\left.\sqrt{{W\p}^2+f}\right|_{z=\tm}\ .
\eeq
This is  the correct result for the holomorphic tension valid for every $m$. The result (\ref{conjecturefirst}), as expected, doesn't keep into account all the corrections (note that the last is evaluated at $m$ and not $\tm$).

The tension of vortices,  contrary to (\ref{BPSmass}), has a non-BPS contribution that is  already present in the simplest case: classical $\N=2$ SQED broken with a superpotential. As noted by many authors \cite{HSZ,VYung}, when the second derivative of the superpotential, ${W\p}\p$, is not zero the tension is not the BPS one. In this case it  is possible to ask when the corrections are small with respect to the holomorphic tension, and also if the confinement is of type \textrm{I} or  type \textrm{II}.

A particularly strong result is obtained when the superpotential is linear. The non-BPS contribution vanishes for every value of $m$ and the holomorphic tension is $4\pi |W\p|$ without quantum corrections.

The paper is organized as follows.  We start our analysis in section \ref{sec:SQED} by reviewing some properties of the vortices that arise in $\N=2$ SQED broken to $\N=1$ by a superpotential. This analysis is completely  classical. We compute the holomorphic tension and  see that the non-BPS contribution  is already present at the classical level. The BPS tension is related to the central charge in the presence of a vortex configuration.

In section \ref{sec:central} we analyze the $\N=2$  $U(N_c)$ gauge theory with $N_f$ flavors, broken to $\N=1$ by a superpotential. The case $N_f<2N_c$ is considered, so that the theory is asymptotically free.   First we analyze the vortices classically where the tension is $T=4\pi W\p(m)$. Then we review the exact results of \cite{CDSW,CSW}  regarding  the chiral ring of the theory. After this we are able to give a first approximation to  the tension in the semiclassical region.

Section \ref{sec:check} is devoted to the computation in the strong coupling regime. The strong coupling is a regime in which the superpotential can be considered a small perturbation to the $\N=2$ theory. So one can  use the exact results on the $\N=2$ low energy action and then  add the effective superpotential.  First we review the Seiberg-Witten curve that describes the low energy $\N=2$ dynamics. Then we recall the results concerning the factorization that relates the $\N=2$ curve to the $\N=1$ curve. Finally we perform the computation of the vortex tension and the result will be (\ref{resultintro}).

In section \ref{sec:limit} we study  the limit in which  the non-BPS contribution can be neglected. We are able to give a condition of validity in the weak coupling regime. Section \ref{sec:conclusion} is devoted to conclusion and discussion.  In \ref{centralcharge} we give some details for the calculation of the central charge in $\N=1$ SQED. In \ref{app:conventions} we relate our conventions with the ones of \cite{CSW}.

\section{Vortices in $\mat{{\cal N} =2}$ SQED \label{sec:SQED}}

The building block of the present work is $\N =2$ SQED. In this section we study this theory  without considering quantum corrections. The $U(1)$ gauge multiplied is composed by the superfields $W_{\alpha}$ and $\Phi$, while the matter superfields are $Q$ of charge $+1$ and $\tQ $ of charge $-1$. $\N=2$ is broken to $\N=1$ by  means of a superpotential that is a holomorphic function of $\Phi$. The Lagrangian is the following:
\bea
\label{classicalSQED}
{\cal L}&=&\int d^2\theta \, \frac{1}{4e^2}W^{\a}W_{\a}+h.c.\\
&&+\int d^2\theta d^2\bar{\theta} \, (\frac{1}{e^2}\Phi\dag\Phi+Q\dag e^VQ+{\widetilde Q}\dag e^{-V}{\widetilde Q})\nonumber\\
&&+\int d^2\theta \, \sqrt{2}({\widetilde Q}\Phi Q-m{\widetilde Q}Q+W(\Phi))+h.c.\ .\nonumber
\eea
The potential for the scalar fields is
\beq
\label{pot}
V=2|(\phi-m)q|^2+2|(\phi-m)\tq|^2+2e^2|\tq q+W\p(\phi)|^2+\frac{e^2}{2}(|q|^2-|\tq|^2)^2\ ,
\eeq
and the vacuum solution is:
\beq
\phi=m\ ,\qquad |q|=|\tq|\ , \qquad \tq q = - W\p(m)\ .
\eeq
The gauge group $U(1)$ is completely broken by the quark condensate, so the theory admits a vortex configuration solution that belongs to the homotopy group $\pi_1(U(1))=\mathbb{Z}$.

\subsection {BPS solution \label{subsec:BPS}}

The homotopy consideration assures us that the vortex solution exists and is stable but in general it is difficult to find it.  Let's start considering the BPS limit:
\beq
\label{BPSansatz}
\phi=m\ , \qquad \tq = -q\dag   \frac{W\p}{|W\p|} \ .
\eeq
Inserting the latter in the potential (\ref{pot}) one obtains
$V(q)=2e^2(|q|^2-|W\p|)^2$. Now the kinetic term of $q$ has a factor  $2$, because it comes from $q$ and $\tq$,  and to restore the proper normalization one has to rescale the field $q \to q/\sqrt{2}$. Thus the Lagrangian becomes the usual one in the BPS limit \cite{Bogomolny}:
\beq
{\cal L}=-\frac{1}{4e^2}F_{\mu\nu}F^{\mu\nu}-(D_{\mu}q)^{\dagger}(D^{\mu}q)-\frac{e^2}{2}(|q|^2-2|W\p|)^2 \ .
\eeq
We orient the vortex in the $\hat{z}$ direction, so the tension is the integral of the energy density in $dx\,dy$, then, using the Bogomoln'y trick, the tension can be written as a sum of quadratic terms plus a boundary term:
\bea
\label{sumquadandbord}
T&=&\int d^2x \,  \frac12 |D_kq+i\epsilon_{kl}D_l q|^2
+(\frac{1}{2e}F_{kl}+\frac{e}{2}(|q|^2-2|W\p|)\epsilon_{kl})^2    \\
&&+ \oint d\vec{x} \cdot ( 2\vec{A}|W\p|-iq^{\dagger}\vec{D}q ) \ .\nonumber
\eea
A vortex of finite energy  must be an element $[n]\in \pi_1(U(1))$
\beq
\label{BPSprofile}
q=  e^{i n\theta}\sqrt{2|W\p|} \, q_n(r)\ , \qquad
A_k =   -n\epsilon_{kl}  \frac{r_l}{r^2} \, f_n(r)\ ,
\eeq
where the profile functions satisfy these boundary conditions: $q_n(\infty),f_n(\infty)=1$ and $q_n(0),f_n(0)=0$. The $q$ field is chosen so that it winds $n$ times at infinity and the $A_k$ field is chosen so that the covariant derivative $D_kq$ vanishes at infinity.   The solution of the equation of motion, that give the profile functions, is the one that minimizes the tension (\ref{sumquadandbord}), so we must put to zero the quadratic terms  and the tension is given by the boundary term, that comes out to be proportional to the flux of the magnetic field:
\beq
\label{BPSsemiclassical}
T=4\pi|\T| \ , \qquad \T=  n W\p(m)  \ .
\eeq
The BPS equations of motion are first order equations obtained by setting to zero the quadratic terms in the tension. Rescaling to a dimensionless length $\rho=e\sqrt{|W\p|}r$ the equations for the profile functions are::
\bea
\label{profileeq}
&& \frac{n}{\rho}\frac{df_{(n)}}{\rho}+{q_{(n)}}^2-1=0\ ,\\
&& \rho \frac{d q_{(n)}}{d\rho}+n(f_{(n)}-1)q_{(n)}=0 \ . \nonumber
\eea
This means that the radius of the vortex is $R_v \sim 1/e\sqrt{|W\p|}$.

Now we discuss an important point regarding the BPS limit (\ref{BPSansatz}). This is only approximate and do not satisfy the complete set of the  equations of motion and the simplest way to see this is to look at the equation of motion for $\phi$:
\beq
\frac{\Box \phi}{e^2}=2(\phi-m)(|q|^2-|\tq|^2)+2e^2 {W\p}\p(\phi)\dag (\tq q+W\p(\phi))\ .
\eeq
With (\ref{BPSansatz}) one would obtain $0=e^2{W\p}\p(m)\dag (|q|^2-|W\p(m)|)$ that is false because inside the radius of the vortex  $|q|^2-|W\p(m)|$ is different from zero. The equation of motion is satisfied only if ${W\p}\p(m)=0$. Naively we can say that the BPS limit, even if not exact, could be a good approximation if $e<<1$. In section \ref{sec:limit} we will give a more detailed analysis of the conditions in which the BPS limit is a good approximation.

\subsection{Central Charge \label{subsec:centralcharge} }

In \ref{centralcharge} we study how the $\N=1$ supersymmetric algebra  is modified by the presence of a vortex configuration that makes possible the existence of a central charge. As explained in \ref{centralcharge}, the central charge implies a BPS bound for the tension of the vortex
\beq
\label{ciaone}
T_{BPS} =   2r \oint d\vec{x} \cdot \vec{A} \ ,
\eeq
where $r$ is the coefficient of the Fayet-Iliopulos term in the Lagrangian.  When this bound is saturated half of the supersymmetries are unbroken.

Now we want to apply the results of \ref{centralcharge} to our theory: $\N=2$ SQED broken to $\N=1$ by a superpotential. A problem immediately arises: the central charge is zero because  there is no FI term.  But here there are two supersymmetries, so the $SU(2)_R$ R-symmetry can be used to get some informations about the central charge. What we are going to do is to perform a $SU(2)_R$ rotation to bring our theory in a form where there is no superpotential but there is a  Fayet-Iliopoulos term \cite{RioandJarah,VYung}. When both the superpotential and the FI term are present the potential is
\beq
V=2e^2|\tq q+W\p(m)|^2+\frac{e^2}{2}(|q|^2-|{\tq}|^2-2r)^2\ , \label{tenthree}
\eeq
and this can be written in a $SU(2)_R$ invariant form:
\bea
&V=e^2\Tr_2\,({q\dag}^{\alpha}q_{\beta}-\frac{1}{2}{\delta^{\alpha}}_{\beta}{q\dag}^{\gamma}q_{\gamma}-\xi_a{(\sigma_a)^{\a}}_{\beta})^2& \ ,\\
&-\xi_1+i\xi_2=W\p(m)\ ,\qquad \xi_3=r \ ,& \nonumber
\eea
where $q^{\alpha}$ is the $SU(2)_R$ doublet $(q,{\tq}\dag )$.  Thus $(\Re W\p(m), \Im W\p(m),r)$ is a triplet of the $SU(2)$ R-symmetry and,
in our case, we may rotate the superpotential away  leaving only an FI term whose coefficient is 
\beq
r=|W\p(m)| \ .
\eeq
Using (\ref{ciaone}) we get the BPS tension
\beq
T_{BPS}= 4\pi |W\p(m)| \ ,
\eeq
and we recover (\ref{BPSsemiclassical}).

\section{Vortices in  $\mat{U(N_c)}$ Theory with $\mat{N_f}$ Flavors:\\ Semiclassical Limit \label{sec:central}}

Here we come to the main subject of this article: the vortices in the $\N =2$ $U(N_c)$ gauge theory with $N_f$ flavors broken  to $\N=1$ by means of  a superpotential. We consider the case $N_f < 2N_c$ so that  the theory is asymptotically free.  To set the conventions we write the Lagrangian of the theory:\footnote{Concerning the generators of the group, we use the $N_c \times N_c$ matrices $T^a\dag=T^a$ with normalization $\Tr_{N_c}\,(T^aT^b)=\delta^{ab}/2$.}
 \bea
\label{QFT}
{\cal L}&=&\int d^2\theta \,\frac{1}{2e^2}\Tr_{N_c}\,(W^{\alpha}W_{\alpha}) +h.c. \\ &&+ \int d^2\theta d^2\bar{\theta}  \,\frac{2}{e^2}\Tr_{N_c}\,(\Phi\dag e^V\Phi e^{-V})+ \int d^2\theta d^2\bar{\theta} \,\sum_{I=1}^{N_f}\, (Q_I\dag e^{V}Q^I+\widetilde{Q}_I e^{-V}\widetilde{Q}^{\dagger I} )\nonumber\\
&& + \int d^2\theta \,\sum_{I=1}^{N_f} \, \sqrt{2}(\widetilde{Q}_I\Phi Q^I - m_I\widetilde{Q}_I Q^I )+\sqrt{2} \Tr_{N_c}\,W(\Phi)  +h.c. \ ,\nonumber
\eea
where
\beq
\label{tree}
W(z)=\sum_{j=0}^{k} \, \frac{g_j}{j+1}z^{j+1}\ ,\qquad W\p(z)=g_k\prod_{j=1}^{k}\, (z-a_j)\ .
\eeq
Here, as in \cite{CSW}, we consider the case in which  all the roots $a_j$ and all the masses $m_I$ are different.

\subsection{Classical analysis}

Now we consider the weak coupling regime (for a detailed analysis see \cite{RioandJarah}).
The diagonal elements of the adjoint field are equal to a flavor mass $m_I$ (the color-flavor locking) or to a root of $W\p$:
\beq
<\phi>=\left(\begin{array}{cccccc}
\ddots&&&&&\\
&m_I {\bf 1}_{r_I}&&&&\\
&&\ddots&&&\\
&&&a_j {\bf 1}_{N_j}&&\\
&&&&\ddots&\\
\end{array}\right)\ ,
\qquad
\sum_{j=1}^{n} \, N_j+\sum_{I=1}^{N_f}\, r_I=N_c\ .
\eeq
The gauge group is classically broken to $ \prod_{j=1}^{n} U(N_j)$, where $n$ is less than or equal to the number of the roots $W\p$ and $r_I$ is less than  or equal to the number of flavors with mass $m_I$. In this case, where all the roots are distinct,  there are only two possibilities: $r_I=0,1$. When $r_I=1$ the flavors $Q^I_a$ and  $\tQ^a_I$ have zero mass and at low energy one has $\N=2$ SQED broken to $\N=1$ by a superpotential, that is the same theory studied in  \ref{classicalSQED}.   Thus the results of section \ref{sec:SQED} can be applied  and so   the theory develops a vortex of tension 
\beq
T=4\pi|W\p(m)| \ .
\eeq

Note that this analysis has been carried out by considering the Lagrangian (\ref{QFT}) classically. When $m>>\Lambda$ we can trust this results as a good approximation because the theory is at weak coupling. In the following we are going to compute the quantum corrections to the tension.  To do this we need first a brief review of the Cachazo-Douglas-Seiberg-Witten solution.

\subsection{CDSW solution \label{exact} }

Here we recall the results of \cite{CDSW,CSW2,CSW} (see \cite{ferretti} for a review), focusing on the points that we need to compute the corrections in subsection \ref{here}. Consider the following operators:
\beq
\label{convenction}
T(z)=\Tr\, \frac{1}{z-\Phi}\ ,\qquad
R(z)=-\frac{1}{16\sqrt{2}\pi^2}\Tr\,\frac{W^{\a}W_{\a}}{z-\Phi} \ ,
\eeq
\beq
M_I(z)=\tQ_I \frac{1}{z-\Phi}Q_I \ .
\eeq
Taking the coefficients of the power expansion in $z$ one obtains all the generators of the chiral ring of the theory. The generalized Konishi anomalies \cite{CDSW,CSW,kenanomaly} provides a solution for the chiral ring. The anomalies  that we need are the following:
\bea
\label{kenanomaly}
&[W\p(z)R(z)]_-= R(z)^2 \ ,\\
&[M_I(z) (z-m_I)]_-= R(z) \ .\nonumber
\eea 
The solution of the first equation is
\beq
\label{R}
2R(z)=W\p(z)-\sqrt{W\p(z)^2+f(z)} \ ,
\eeq
where $f(z)$ is a polynomial of degree $k-1$ that depends on the vacuum. By (\ref{R}) we are naturally led to consider the Riemann surface $\Sigma_{\N=1}$ defined by the equation
\beq
\label{SigmaN=1}
y^2=W\p(z)^2+f(z) \ .
\eeq
This is a double sheeted cover of the complex plane on which $R(z)$ is uniquely defined. We call $q_I$ and $\tq_I$  the two points of $\Sigma_{\N=1}$ with the same  coordinate $z=m_I$.  When $r_I=0$ there is no color-flavor locking and $M_I(z)$  must be regular in $q_I$:
\beq
M_I(z)=\frac{R(z)}{z-m_I}-\frac{R(q_I)}{z-m_I} \ .
\eeq
When $r_I=1$ one can find the solution by continuously deforming the theory, in such a way that the pole passes from the second to the first sheet. The solution is:
\beq
\label{M}
M_I(z)=\frac{R(z)}{z-m_I}-\frac{W\p(m_I)-R(q_I)}{z-m_I} \ .
\eeq

In the semiclassical limit, where $f(z) \to 0$, the cuts of $\Sigma_{\N=1}$ are closed and the Riemann surface becomes simply the Riemann sphere $\hat{\mathbb{C}}$ with punctures at $a_j$ and $m_I$. The gaugino condensate is $R(z)=0$, as it should be, and the quark condensate is
\beq
M_I(z)=-\frac{W\p(m_I)}{z-m_I}\ .
\eeq

\subsection{The holomorphic tension: a first approximation  \label{here}}

In the semiclassical limit  ($m>>\Lambda$)  the quarks $Q_I$ and $\tQ_I$ are the low energy degrees of freedom that, upon breaking to $\N=1$, condense and create the vortex.  So we expect that the holomorphic tension is given by their condensate as in (\ref{classicalSQED})
\beq
\label{aafirst}
\T_{I}=- \tQ_I Q_I \ .
\eeq
This condensate belongs to the chiral ring and can be computed using the results of subsection \ref{exact}. $\tQ_I Q_I$ is given by the $1/z$ pole of the generator $M_I(z)$
\beq
\label{bbbbbbbbbbbbbbbbbb}
 \tQ_I Q_I=\frac{1}{2\pi i} \oint_{\infty} M_I(z) dz = -\frac{1}{2} \left.(W\p + \sqrt{{W\p}^2+f}) \right|_{z=m_I} \ ,
\eeq

This result is only a first approximation to the holomorphic tension because  (\ref{aafirst}) would be exact only if the low energy superpotential was:
\beq
\W_{low}=
\sqrt{2} (\tQ_I (\Phi-m_I) Q_I + W_{eff}(\Phi)) \ .
\eeq
%ith  
%\beq
%W_{eff}(z)= \int dz \sqrt{{W\p}^2(z)+f(z)} \ .
%\eeq
On the other hand, in the low energy Lagrangian there could also be terms like
$\Tr_{N_c} \, \Phi^{\alpha}$, $\tQ_I \Phi ^{\beta}  Q_I$ or products of them.
In particular terms such as  $\tQ_I  P(\Phi)  Q_I$, where $P(z)$ is a polynomial,  will have the effect of shifting the mass $m_I$.  The right result  will be  computed in the next section using the factorized curves in the strong coupling regime (see (\ref{strongtension})).

\section{Strong Coupling  Computation of $\mathbf{\T}$ \label{sec:check}}

Here we compute the holomorphic tension  at strong coupling, where $W$ is a small perturbation of the $\N=2$ theory. In this regime we can take the low energy $\N=2$ theory and add the effective superpotential generated by $W$.

\subsection{Low energy $\mat{\N=2}$}

Here we recall the results about the low energy dynamics of the $\N=2$ theory (see \cite{hananyoz} for the conventions and \cite{alvarez} for a review). 

First we consider $SU(N_c)$ with $N_f$ flavors. At low energy one has $N_c-1$ $U(1)$ gauge multiplets and we call their scalar components $a_i$, where $i=1,\dots,N_c-1$. The moduli space is a $N_c-1$ dimensional complex manifold ${\cal M}_{SU(N_c)}$, parametrized by the gauge invariant coordinates
\beq
u_j=\frac{1}{j}\langle\Tr\, \phi^j\rangle\ , \qquad  j=2,\dots,N_c \ .
\eeq 
The informations concerning the low energy dynamics are encoded in the  Riemann surface $\Sigma_{\N=2}$ defined by
\beq
\label{SigmaN=2}
{y}^2=P_{N_c}(z)^2-\Lambda^{2N_c-N_f}\prod_{I=1}^{N_f}(z-m_I)\ ,
\eeq
where $P_{N_c}(z)=\det(z-\phi)$ can be written in power series of $z$
\beq
 P_{N_c}(z)=\sum_{k=0}^{N_c} \, s_k z^{N_c-k} \ , 
\eeq
\beq
 s_0=1\ , \quad s_1=0 \ , \quad s_k=(-)^k \sum_{i_1 <\dots < i_k} \, \phi_{i_1} \dots \phi_{i_k} \ .
\eeq
Being $\Sigma_{\N=2}$  a genus $N_c-1$ Riemann surface, we can choose $N_c-1$ independents holomorphic differentials:\footnote{The normalization will be fixed imposing the correct semiclassical result.}
\beq
\lambda_j \propto \frac{ z^{N_c-j} dz }{y}\ , \qquad j=2,\dots, N_c \ .
\eeq
Each $a_i$ corresponds to an $\alpha_i$ cycle on $\Sigma_{\N=2}$, while its dual $a_{Dj}$ corresponds to a $\beta_j$ cycle chosen in such a way that the intersection is $<\alpha_i,\beta_j>=\delta_{ij}$. The solution is given by the period integrals
\beq
\label{Solution}
\frac{\de a_i}{\de s_j}=\oint_{\alpha_i}\lambda_j\ ,\qquad
\frac{\de a_{Di}}{\de s_j}=\oint_{\beta_i}\lambda_j\ .
\eeq
 The relation between $u_j$ and $s_k$ will be important for us because the solution (\ref{Solution}) gives $\de a_i / \de s_j$ but, to calculate the tension, we will need $\de a_i / \de u_j$.  These relations can be encoded in  a single one \cite{CSW2}:
\beq
\label{nicerelation}
P_{N_c}(z)= z^{N_c} \exp{\left(-\sum_{j=1}^{\infty} \, \frac{u_j}{z^j} \right)_+} \ ,
\eeq 
where by $(\;)_+$ we mean that we discard the negative power expansion.

Now we study the $U(N_c)$ theory with $N_f$ flavors and we are going to see that the solution can be easily incorporated in the previous ones, with few modifications. The low energy theory has one more $U(1)$ factor that comes  from the decomposition $U(N_c)=U(1)\times SU(N_c)$ and we denote its scalar component with $a_{N_c}$. This factor has no strong dynamics: in the $N_f=0$ case it is completely free, while in the $N_f\neq0$ case it is infrared free. The moduli space ${\cal M}_{U(N_c)}$ has one dimension more and is parametrized by
\beq
 u_j=\frac{1}{j} \langle\Tr\, \phi^j\rangle \ , \qquad  j=1,\dots,N_c \ .
\eeq 
The Riemann surface is the same given  in (\ref{SigmaN=2}), but here $\phi$ can have non zero trace and $\Sigma_{\N=2}$ depends also on the modulus $u_1$. To complete our task we must find the cycle $\alpha_{N_c}$ that corresponds to $a_{N_c}$ and the differential $\lambda_1$ that corresponds to $s_1$. The cycle $\alpha_{N_c}$ is the one that encircles all the cuts in the $z$ plane. Note that this is a trivial cycle and only a meromorphic differential can be different from zero when it is  integrated around it. The differential that corresponds to  $s_1=-u_1$ is
\beq
\lambda_1 \propto \frac{ z^{N_c-1} dz}{y} 
\eeq
and  is meromorphic because it has a pole at $\infty$.   With these  modifications the solution is encoded in (\ref{Solution}).

\subsection{Breaking to $\mat{\N=1}$}

Now we break $\N=2$ to $\N=1$ by a superpotential.  This breaking leaves  only a discrete number of vacua.  If the low energy is $U(1)^n$, than $N_c-n$ gauge factors are broken by the condensate of a  charged field. These charged fields must be massless in the $\N=2$ theory and  so the Riemann surface must have $N_c-n$ degenerate branch cuts
\beq
\label{factorone}
{y}^2=P_{N_c}(z)^2-\Lambda^{2N_c-N_f}\prod_{I=1}^{N_f}(z-m_I)=
F_{2n}(z)H_{N_c-n}(z)^2 \ .
\eeq
The connection with the $\N=1 $ curve (\ref{SigmaN=1}) is given by the following factorization \cite{CSW2,CSW,deBoer,largeN,Ook}
\beq
\label{factortwo}
{y_{m}}^2=W\p(z)^2+f(z)={g_{k}}^2 F_{2n}(z) Q_{k-n}(z)^2 \ .
\eeq
The above  factorization provides  all the informations we need: from the knowledge of $W$ one can obtain the vacua that survive and the correspondent $\N=1$ curve. As in the classical case,   two conditions must be satisfied: $n\leq k$ and $n\leq N_c$. When  $k=n$ the factorization is simply
\beq
\label{factorthree}
{y}^2=P_{N_c}(z)^2-\Lambda^{2N_c-N_f}\prod_{I=1}^{N_f}(z-m_I)=
\frac{1}{{g_{k}}^2}({W\p}^2(z)+f(z))H_{N_c-n}(z)^2 \ .
\eeq

Consider the case where $N_c-n$  roots of (\ref{SigmaN=2}) collide and, as we said before, $N_c-n$ of the $U(1)$ gauge factors will have massless charged field that we call $Q_r,\tQ_r$, with $r=1,\dots,N_c-n$.  The low energy superpotential is the $\N=2$ one plus a holomorphic function of the chiral superfields $A_i$ \beq
\label{lowsup}
\W_{low}=\sqrt{2}(\sum_{r=1}^{N_c-n}\, \tQ_r A_r Q_r + W_{eff}(A_1,\dots,A_{N_c})) \ .
\eeq
By means of holomorphic arguments \cite{SW1} one can show that, when the tree superpotential is (\ref{tree}), the effective superpotential is 
\beq
W_{eff}=\sum_{j=1}^{k} \, g_j u_{j+1}(A_1,\dots,A_{N_c}) \ ,
\eeq
where the $u_{j+1}(A_1,\dots,A_{N_c})$ are given implicitly by the solution (\ref{Solution}).
For our proof  we need to consider only the $F_A$ terms of the potential:
\beq
F_{A_r}=2 {e_r}^2 |\tq_r q_r + \frac{\de W_{eff}}{\de a_r}|^2\ , \qquad
r=1,\dots,N_c-n \ , 
\eeq
\beq
F_{A_s}=2 {e_s}^2 |\frac{\de W_{eff}}{\de a_s}|^2\ , \qquad s=N_c-n+1,\dots,N_c \ . 
\eeq
The first one  gives  the holomorphic tension of the $r$-vortex, while the second gives a stationary condition:
\beq
\label{formulas}
\T_r=-\tq_r q_r=\frac{\de W_{eff}}{\de a_r} \ , 
\eeq
\beq
\label{formulas2} 
0=\frac{W_{eff}}{\de a_s} \ .
\eeq

\subsection{Computation of $\mathbf{\T}$}

We are now ready to perform the computation of the holomorphic tension. We will start by the  simplest case, then we will consider step by step  more general cases.

\subsubsection*{ $n=k=N_c-1$ }

\begin{figure}[ht]
\begin{center}
\leavevmode
\epsfxsize 12   cm
\epsffile{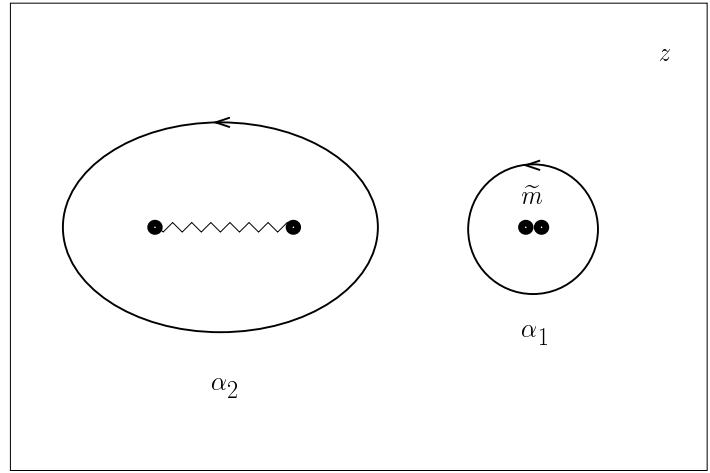}    
\end{center} 
\caption{Cycles in $U(2)$ theory.}
\label{cycleU(2)} 
\end{figure}

The simplest case of this category is $N_c=2$ with $k=n=1$ where the superpotential is
\beq
W(z)=g_0z+\frac{g_1}{2}z^2\ .
\eeq
The Riemann surface $\Sigma_{\N=2}$ has two cuts (see figure \ref{cycleU(2)}), and the first one is shrank  to a point $z=\tm$.  We denote by $\alpha_1$ and  $\alpha_2$ the cycles encircling the two cuts. 
The factorization (\ref{factorthree}) gives 
\beq
\label{factorU(2)}
\Sigma_{\N=2}: \qquad y^2=\frac{1}{{g_{1}}^2}({W\p}^2+f)(z-\tm)^2 \ ,
\eeq
while the integrals around the cycle $\a_1$ become simply the residues around the point $\tm$, for example
\beq
\label{residue1}
\frac{1}{2\pi i} \oint_{\a_1} \frac{dz}{y}=\frac{g_1}{\sqrt{{W\p}^2+f}} \ .
\eeq
In this case the relation between $s$ and $u$  is:
\beq
s_1=-u_1 \ , \qquad  s_2 = - u_2 +\frac{{u_1}^2}{2} \ .
\eeq 
For our proof  we will need  to calculate only
\beq
\frac{\de a_1}{\de u_2}=\frac{\de a_1}{\de s_1}\frac{\de s_1}{\de u_2}+\frac{\de a_1}{\de s_2}\frac{\de s_2}{\de u_2}=-\frac{\de a_1}{\de s_2} .
\eeq
First, we observe that the solution (\ref{Solution}) and the residue (\ref{residue1}) give:\footnote{The proper normalization of the holomorphic differential to reproduse the correct semiclassical result is $-1/2\pi i$}
\beq
\label{op}
\frac{\de a_1}{\de s_2}=-\frac{g_1}{\sqrt{{W\p}^2+f}} \ .
\eeq
Then, writing the equations (\ref{formulas}) and (\ref{formulas2}) in a matrix form
and left  multiplying  by the inverse matrix, we get
\beq
\label{matrice}
\left(
\begin{array}{c}
g_0\\
g_1\\
\end{array}
\right)
=
\left(
\begin{array}{cc}
\de a_1/\de u_1&\de a_2/\de u_1\\
\de a_1/\de u_2&\de a_2/\de u_2\\
\end{array}
\right)
\left(
\begin{array}{c}
\T\\
0\\
\end{array}
\right)\ .
\eeq
The simple passage of multiplying by the inverse matrix has simplified a lot our work because now (\ref{matrice}) is expressed as a function of  $\de a_i/\de u_j$, known directly through (\ref{Solution}).  Furthermore only $\de a_1 / \de u_{1,2}$, the ones obtained by an integral around the collided branch,  are important because the others are multiplied by zero.
Actually by (\ref{matrice}) we need only the second equation
\beq
\label{lastone}
g_1= \T  \frac{\de a_1}{ \de u_2} \ ,
\eeq
that, using (\ref{op}), gives the holomorphic tension:
\beq
\label{strongtension}
\T=\left.\sqrt{{W\p}^2+f}\right|_{z=\tm}\ .
\eeq

Let us see what happens in the general case $n=k=N_c-1$.  We can write a matrix equation like (\ref{matrice}) where the couplings vector is
\beq
{\bf g}=(g_0, \dots,  g_{N_c-1}) 
\eeq
and the tension vector is 
\beq
{\bf T}=(\T,  0 , \dots,  0) \ .
\eeq
With these conventions the matrix equation that generalizes (\ref{matrice}) becomes:
\beq
\label{generalmatrice}
{\bf g}= \mathbf{\frac{\de a}{\de  u}  T} \ .
\eeq
As in the simplest case examined before, where we needed only the second equation of (\ref{matrice}), now  we need only the last one among the equations contained in the matrix relation  (\ref{generalmatrice}):
\beq
g_{N_c-1}=\T \frac{\de a_1}{\de u_{N_c}} \ .
\eeq
The derivative can be easily calculated
\beq
\frac{\de a_1}{\de u_{N_c}}=-\frac{\de a_1}{\de s_{N_c}}=\frac{g_{N_c-1}}{\sqrt{{W\p}^2+f}} 
\eeq
and one still gets (\ref{strongtension}) for the holomorphic tension.

\subsubsection*{ $ n=k<N_c-1 $ }

The next step consists in letting more flavors to be locked. We still consider $k=n$, in such a way that the factorization is the simplest one, given in (\ref{factorthree}). 

First we give the proof for the simplest example of this kind i.e. $U(3)$ with two flavors of mass $m_1$ and $m_2$. The superpotential is 
\beq
W(z)=g_0z+\frac{g_1}{2}z^2\ ,
\eeq
while the factorization gives
\beq
\Sigma_{\N=2}: \qquad  y^2=\frac{1}{{g_{1}}^2}({W\p}^2+f)(z-\tm_1)^2(z-\tm_2)^2\ .
\eeq
The relation between $s$ and $u$ in this case is:
\beq
s_1=-u_1 \ , \qquad  s_2 = - u_2 +\frac{{u_1}^2}{2} \ , \qquad
s_3=-u_3+u_1 u_2 - \frac{{u_1}^3}{6} \ .  
\eeq 
By using the same trick explained before, we write equations (\ref{formulas}) and (\ref{formulas2}) in a matrix form and  multiply by the inverse, obtaining
%\beq
%\label{short}
%{\bf g}_i=  \frac{\de a_j}{\de u_i} {\bf T}_j
%\eeq
\beq
\label{matricedue}
\left(
\begin{array}{c}
g_0\\
g_1\\
0\\
\end{array}
\right)
=
\left(
\begin{array}{ccc}
\de a_1/\de u_1&\dots&\de a_3/\de u_1\\
\vdots&&\vdots\\
\de a_1/\de u_3&\dots&\de a_3/\de u_3\\
\end{array}
\right)
\left(
\begin{array}{c}
\T_1\\
\T_2\\
0\\
\end{array}
\right)\ .
\eeq
Now, as in the previous case,  we need to  calculate only the  residues around $\tm_1$ and $\tm_2$.
The last equation of (\ref{matricedue}) is enough for our proof: 
\beq
\label{sistem1}
%1&=&   \T_1\frac{ m_1}{\left.\sqrt{{W\p}^2+f}\right|_{m_1} (m_1-m_2)   } + \T_2 \frac{ m_2}{\left.\sqrt{{W\p}^2+f}\right|_{m_2}(m_2-m_1)}  \ ,\\
0 =   \T_1\frac{1}{\left.\sqrt{{W\p}^2+f}\right|_{\tm_1} (\tm_1-\tm_2)   } +   \T_2\frac{1}{\left.\sqrt{{W\p}^2+f}\right|_{\tm_2}(\tm_2-\tm_1)} \ .
%\frac{g_0}{g_1}&=& T_1 T_1\frac{m_1^2}{\left.\sqrt{{W\p}^2+f}\right|_{m_1} (m_1-m_2)   } + T_2  \frac{m_2^2}{\left.\sqrt{{W\p}^2+f}\right|_{m_2}(m_2-m_1)}   \nonumber
\eeq
If we impose that $\T_1$ does not depend on $\tm_2$ and $\T_2$ does not depend on $\tm_1$
\footnote{We mean that $\T_I$ doesn't depend on $\tm_{J \neq I}$ if we vary $\tm_J$ keeping fixed $W(z)$ and $f(z)$.}, the solution is unique:
\beq
\label{solutionone}
\T_1=\left.\sqrt{{W\p}^2+f}\right|_{\tm_1} \ ,\qquad \T_2=\left.\sqrt{{W\p}^2+f}\right|_{\tm_2}\ .
\eeq

The requirement  that $\T_I$ depends only on its mass $\tm_I$ has  brought a great simplification because only the last of the matrix equation is necessary to find the solution.    Actually in this simple case one can verify that the requirement is indeed true  by using also the second equation of (\ref{matricedue}). Using
\beq
\frac{\de a}{\de u_2}= -\frac{\de a}{\de s_2} + u_1 \frac{\de a}{\de s_3} \ ,
\eeq
the second equation of (\ref{matricedue}) leads to another independent equation
\beq
1=   \T_1\frac{ \tm_1}{\left.\sqrt{{W\p}^2+f}\right|_{\tm_1} (\tm_1-\tm_2)   } + \T_2 \frac{ \tm_2}{\left.\sqrt{{W\p}^2+f}\right|_{\tm_2}(\tm_2-\tm_1)}  \ .
\eeq
This equation together with (\ref{sistem1}) are enough to establish the solution (\ref{solutionone}) and in particular to verify that $\T_I$ depends only on $\tm_I$.

We now consider the general case $n=k<N_c-1$. The factorization of the curve gives
\beq
\Sigma_{\N=2}: \qquad  y^2=\frac{1}{{g_n}^2}({W\p}^2+f)(z-\tm_1)^2\dots(z-\tm_{N_c-n})^2\ ,
\eeq
while the vectors of the coupling and  of the tension are:
\bea
{\bf g}&=&(g_0 , \dots , g_{n} , 0 , \dots ,0) \ , \\
{\bf T}&=&(\T_1 , \dots , \T_{N_c-n},  0 , \dots , 0) \ . \nonumber
\eea
The last $N_c-n$ equations of (\ref{generalmatrice}) are:
\bea
\label{lungheequazioni}
1&=& \sum_J \, \T_J   \frac{\tm_{J}^{\phantom{J} N_c-n-1}}{ \left.\sqrt{{W\p}^2+f}\right|_{\tm_J}\prod_{I \neq J} (\tm_J-\tm_I)}  \ , \\ 
0&=& \sum_J \, \T_J   \frac{\tm_{J}^{\phantom{J}N_c-n-2}}{ \left.\sqrt{{W\p}^2+f}\right|_{\tm_J}\prod_{I \neq J} (\tm_J-\tm_I)}  \ , \nonumber \\ 
&\vdots &  \nonumber \\
0&=& \sum_J \, \T_J   \frac{1}{ \left.\sqrt{{W\p}^2+f}\right|_{\tm_J}\prod_{I \neq J} (\tm_J-\tm_I)}  \ . \nonumber 
\eea
If we put the desired result $\T_J=\left.\sqrt{{W\p}^2+f}\right|_{\tm_J}$, we obtain the following terms with $r=1 \dots N_c - n  $:
\beq
\label{tech}
\sum_J \, \frac{\tm_J^{\phantom{J}r}}{ \prod_{I \neq J} (\tm_J-\tm_I)} = \frac{\sum_J \,  (-)^J \tm_J^{\phantom{J}r} \prod_{I <  K; \, I,K \neq J} (\tm_I-\tm_K)}{\prod_{I <  K} (\tm_I-\tm_K)} \ .
\eeq
It's easy to see that the numerator is a multiple of the denominator because  if  some $\tm_I$ are equal to some of the $\tm_K$ the numerator vanishes. On the other hand if $r < N_c-n$ the  proportionality constant must vanish  because its power is less that the power of the denominator. When $r=N_c-n$, the fraction is equal to one. Summarizing the results we obtain:  
\beq
\label{techsolution}
\sum_J \, \frac{\tm_{J}^{\phantom{J}r}}{ \prod_{I \neq J} (\tm_J-\tm_I)} = 
\left\{
\begin{array}{c}
0  \qquad r < N_c-n  \\
1 \qquad r=N_c-n  \\ 
\end{array}\right.
\eeq
and the equations (\ref{lungheequazioni}) are satisfied.

\subsubsection*{ $n<k\leq N_c$}

Here the case $n<k$ is considered with  the factorization  given by (\ref{factorone}), (\ref{factortwo}). The simplest example of this category is $U(3)$ with two flavors and the following superpotential:
\beq
W(z)=g_0z+\frac{g_1}{2}z^2+\frac{g_2}{3}z^3\ .
\eeq
If one denotes $Q_{k-n}(z)=z-\gamma$ in (\ref{factortwo}), then the factorization gives 
\beq
\Sigma_{\N=2}: \qquad
y^2=\frac{1}{{g_{2}}^2}({W\p}^2+f)\frac{(z-\tm_1)^2(z-\tm_2)^2}{(z-\gamma)^2} \ .
\eeq
The matrix equation can be written as
\beq
\left(
\begin{array}{c}
g_0\\
g_1\\
g_2\\
\end{array}
\right)
=
\left(
\begin{array}{ccc}
\de a_1/\de u_1&\dots&\de a_3/\de u_1\\
\vdots&&\vdots\\
\de a_1/\de u_3&\dots&\de a_3/\de u_3\\
\end{array}
\right)
\left(
\begin{array}{c}
\T_1\\
\T_2\\
0\\
\end{array}
\right)
\eeq
and the last equation is
\beq
1=  \T_1\frac{\tm_1-\gamma}{\left.\sqrt{{W\p}^2+f}\right|_{\tm_1} (\tm_1-\tm_2)   } +  \T_2\frac{\tm_2-\gamma}{\left.\sqrt{{W\p}^2+f}\right|_{\tm_2}(\tm_2-\tm_1)} \ . \\
\eeq
If we impose that $\T_1$ depends only on $m_1$ and that $\T_2$ depends only on $m_2$, the unique solution of this equation is again (\ref{solutionone}).

In the general case $n<k\leq N_c$ the factorization is
\beq
\Sigma_{\N=2}: \qquad  y^2=\frac{1}{{g_{k}}^2}({W\p}^2+f)\frac{(z-\tm_1)^2\dots(z-\tm_{N_c-n})^2}{(z-\gamma_1)^2\dots(z-\gamma_{k-n})^2}       
\eeq
while the coupling and the tension vectors are:
\bea
{\bf g}&=&(g_0 , \dots , g_{k} , 0 , \dots , 0) \ , \\
{\bf T}&=&(\T_1 , \dots , \T_{N_c-n},  0 , \dots , 0) \ . \nonumber
\eea
We use only the last one of the matrix equation and the condition that $\T_I$ depends only on $\tm_I$. Two cases must be distinguished: the case  $k<N_c$ where the last equation is
\beq
0= \sum_J \, \T_J   \frac{\prod_{q=1}^{k-n} (\tm_J-\gamma_q)}{ \left.\sqrt{{W\p}^2+f}\right|_{\tm_J}\prod_{I \neq J} (\tm_J-\tm_I)} \ ,
\eeq
and the case  $k=N_c$, where it is 
\beq
1= \sum_J \, \T_J   \frac{\prod_{q=1}^{N_c-n} (\tm_J-\gamma_q)}{ \left.\sqrt{{W\p}^2+f}\right|_{\tm_J}\prod_{I \neq J} (\tm_J-\tm_I)} \ .
\eeq
Using (\ref{tech}) one can show that the solution is $\T_J=\left.\sqrt{{W\p}^2+f}\right|_{\tm_J}$.

\subsubsection*{ General case: $n \leq k$ and $n \leq N_c$}

Let us finally consider the most general case: $n \leq k$ and $n \leq N_c$.
What is new here is that $k$ can be greater than $N_c$ and so to compute the tension we should evaluate also the derivatives $\de u_{k>N_c} / \de a$. Only the first $N_c$ of the $u_k$ are  independents, and  the expansion in negative powers of (\ref{nicerelation}) gives the classical relations to obtain the $u_{k>N_c}$'s as functions of the previous ones. Some of these relations gets quantum corrections, so the brute force computation is difficult.

The assumption that $\T_I$ doesen't depends on $\tm_{J\neq I}$ helps us to overcome this problem. Keeping $W(z)$ and $f(z)$ fixed we can add a color locked flavor of mass $\tm_J$ bringing it from infinity. This procedure do not change $k$, $n$ and, by our assumption, neither $\T_I$. The effect of this change is to obtain a theory with one more color and one more flavor
 \beq
N_c \ , N_f \longrightarrow N_c+1 \ , N_f+1 \ .
\eeq
We can repeat this procedure untill the new $N_c$ is greater than $k$, and so the tension can be calculated as in the previous cases.

\subsubsection*{ $\T_I$ doesn't depend on $\tm_{J\neq I}$}

Our check was based on an assumption still not completely proved: the tension $\T_I$ do not change if we move $\tm_{J\neq I}$ and  keep fixed $W(z)$ and $f(z)$. We stress that this information should be contained in the full set of equations, but only for the simplest cases  we were able to give a complete computation without using it.

Here, based on the classical limit and  holomorphy, we give a valid argument in favor of this assumption. Holomorphy tells that, if a function doesn't depends on some parameter in some region, it doesn't depend on that parameter globally.  In the classical limit the holomorphic tension is $W\p(m_I)$. Suppose that there are holomorphic corrections that depends on $m_{J \neq I}$, they should be of two types: a positive power or a negative power of $m_{J \neq I}$. The first case, as example 
\beq
\frac{\Lambda m_{J\neq I}}{m_I^{\phantom{I}2}} \ ,
\eeq
 can be immediately excluded. If we bring $m_{J \neq I}$ to infinity this correction grows but the flavor should decouple from the theory. Also the second case, as example
\beq
\frac{\Lambda}{m_{J\neq I}} \ ,
\eeq
 can be excluded. This correction should grows if we bring  $m_{J \neq I}$ to zero and at some point dominate the classical tension  $W\p(m_I)$, but if we simultaneously  bring $m_I$ to infinity the classical tension should dominate.  
This contradictions exclude the possibility of holomorphic corrections to the $I$-tension that depends on $m_{J\neq I}$.

\section{Limit of Validity \label{sec:limit}}

The tension, as we said in subsection \ref{subsec:BPS}, has a non-BPS contribution that we are not able to compute. Here  we consider in more detail this contribution  and  the limit in which it can be neglected because small with respect to the BPS tension.

\subsection{Classical and quantum SQED}

 First consider  SQED at a classical level.  We said that the BPS vortex  (\ref{BPSansatz}) do not satisfy the complete set of the equation of motion, so we write the tension from the Lagrangian (\ref{classicalSQED})  without making any approximation
\bea
\label {nonBPS}
T & =& \int d^2 x \,  \frac{1}{4e^2}F_{kl}F_{kl}+\frac{1}{e^2}\de_k\phi\dag\de_k\phi+(D_{k}q)^{\dagger}(D_{k}q)+(D_{k}\tq)^{\dagger}(D_{k} \tq) \\
&&+ 2|(\phi-m)q|^2+2|(\phi-m)\tq|^2+2e^2|\tq q+W\p(\phi)|^2+\frac{e^2}{2}(|q|^2-|\tq|^2)^2 \ .\nonumber
\eea
 We want to compute the first order correction to the BPS tension and, comparing it to the BPS tension,  we  get the condition in which  the non-BPS contribution can be neglected.  The stationary equations deriving from (\ref{nonBPS}) are:
\bea
\label{nonBPSequation}
&\triangle \phi/e^2=2(\phi-m)(|q|^2-|\tq|^2)+2e^2 {W\p}\p(\phi)\dag (\tq q+W\p(\phi))\ ,& \\
&\triangle q=\dots \ ,\qquad \triangle \tq =\dots \ ,\qquad \de_kF_{kl} =\dots\ .& \nonumber
\eea
To compute the first order correction we use the following technique. First we put in the right member of the equations of motion the BPS solution (\ref{BPSansatz}) and (\ref{BPSprofile}) and  then we solve obtaining the  solution corrected to the first order. The equations in the second row of (\ref{nonBPSequation}) need no corrections, thus the only correction is given by $\phi=m+\phi_{(1)}$ and the equation in the first row
\beq
\label{firstcorrection}
\triangle \phi_{(1)} = 2e^4 {W\p}\p(m)\dag (\tq q+W\p(m)) \ .
\eeq
From this equation we are able to give an estimation for $\phi_{(1)}$. Outside the radius of the vortex $R_v$, $\phi_{(1)}$ is zero, while inside  $\phi_{(1)} \sim e^4 {W\p}\p W\p {R_{v}}^2$ (remember that from (\ref{profileeq}) $R_v \sim 1/e\sqrt{|W\p|}$).
The first order  correction to the tension comes from three pieces: the first is the kinetic term of $\phi$
\beq
\label{onecor}
\int d^2 x \, \frac{1}{e^2} \de_k \phi_{(1)}\dag \de_k\phi_{(1)} \sim e^6 {{W\p}\p}^2 {W\p}^2 {R_{v}}^4 \sim e^2 {{W\p}\p}^2 \ ,
\eeq
 the second is the sum of the $F_q$ and $F_{\tq}$ terms
\beq
\label{twocor}
\int d^2x \,  2|\phi_{(1)}q|^2+2|\phi_{(1)}\tq|^2  \sim e^8 {{W\p}\p}^2 {W\p}^3 {R_{v}}^6 \sim e^2 {{W\p}\p}^2 \ .
\eeq
and the last  is the deformation of the $F_{\phi}$ term
\beq
\label{threecor}
\int d^2x \,  2e^2 \frac{\de}{\de \phi} |\tq q+W\p(\phi)|^2 \phi_{(1)}\sim e^6 {{W\p}\p}^2 {W\p}^2 {R_{v}}^4 \sim e^2 {{W\p}\p}^2 \ .
\eeq
All these three corrections are of the same order and so the holomorphic tension is a good approximation to the real tension if the following condition is satisfied:
\beq
\label{goodone}
\frac{e^2{{W\p}\p}^2}{{W\p}} << 1\ ,
\eeq
where we don't write the modulus for convenience. 
See \cite{Hou} for a numerical computation or the first order correction that we have estimate above.

Now we consider quantum corrections to SQED.  Being this theory   infrared free,  the coupling constant at low energy goes like
\beq
\label{oneone}
\frac{1}{e^2} \sim \log{\frac{\Lambda_{U(1)}}{\mu}}, \qquad \mu<<\Lambda_{U(1)} \ .
\eeq
The condition (\ref{goodone}) becomes:
\beq
\label{goodtwo}
\frac{{{W\p}\p}^2} {\log{\left(\Lambda_{U(1)} / \sqrt{W\p}  \right)} {W\p}} << 1 \ ,
\eeq
where $\sqrt{W\p}$ is the energy scale of the $U(1)$ breaking.

\subsection{ $\mat{U(N_c)}$ theory with $\mat{N_f}$ flavors: semiclassical limit }

\begin{figure}[ht]
\begin{center}
\leavevmode
\epsfxsize 15   cm
\epsffile{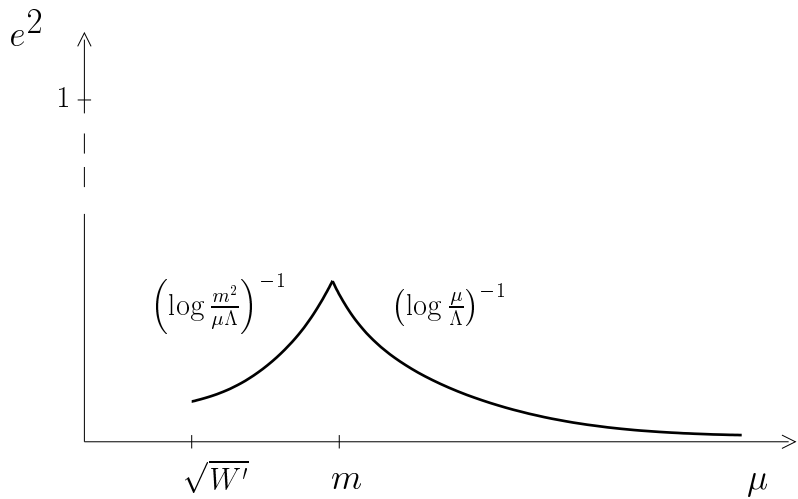}    
\end{center} 
\caption{RG flow of $e^2$ at weak coupling ($m>>\Lambda$).}
\label{RGflow} 
\end{figure}

Now we embed the $U(1)$ theory in the asymptotically free (AF) theory $U(N_c)$ with $N_f$ flavors (see figure \ref{RGflow}). 
The AF theory has a dinamical scale $\Lambda$, and the coupling constant at high energy  goes like
\beq
\label{twotwo}
\frac{1}{e^2} \sim \log{\frac{\mu}{\Lambda}}, \qquad \mu>>\Lambda \ .
\eeq
To find $\Lambda_{U(1)}$ one has to match (\ref{oneone}) with (\ref{twotwo}) when $\mu =m$. Note that this is reliable only if $m>>\Lambda$ and so  in the weak coupling regime. With this matching we obtain $\Lambda_{U(1)} \sim m^2/ \Lambda$ and  the condition (\ref{goodtwo})  becomes:
\beq
\label{goodthree}
\frac{{{W\p}\p}^2}{\log{\left( m^2 / \Lambda \sqrt{W\p} \right)} {W\p}} << 1 \ .
\eeq

\subsection{ $\mat{U(N_c)}$ theory with $\mat{N_f}$ flavors: strong coupling}

Finally we come to the strong coupling limit. The theory is described by a dual $U(1)$. The dual-quark condensate breaks the $U(1)$ at a scale lower than $\Lambda_{U(1)}$. The condition (\ref{goodone}) under which the non-BPS correction is small  becomes
\beq
\label{g}
\frac{e^2(\mu) \, {{W_{eff}\p}\p}^2}{{W_{eff}\p}} << 1\ ,
\eeq
where we have considered $W_{eff}$ that enters in (\ref{lowsup}). The energy scale $\mu$ of the $U(1)$ breking is roughly $\sqrt{W_{eff}\p}$. We argue that a region of parameters exists where the condition (\ref{g}) is satisfied. To find it we multiply the tree level superpotential by a constant $\epsilon$:
\beq
\epsilon W(z)\ , \qquad 0\leq \epsilon \leq 1 \ .
\eeq 
If we send $\epsilon \to 0$, the BPS tension goes to zero like $ \epsilon$ while the non-BPS correction goes to zero more quickly. In fact ${{W_{eff}\p}\p}^2$ brings a factor $\epsilon^2$ and $e^2(\mu)$ vanishes logarithmically with $\epsilon$. Thus for sufficient little $\epsilon$ our vortices are almost BPS.\footnote{If ${{W_{eff}\p}\p}^2 / W_{eff}\p \leq 1$, the strong coupling region, where $e^2(\mu)$ is small, is enough for (\ref{g}) to be valid.}

\subsection{Type \textrm{I} or type \textrm{II}?}

The BPS tension scales linearly with the winding number $n$
\beq
T_{BPS}=4\pi |n\T| \ .
\eeq
What about the real tension? Considering (\ref{nonBPS}), we can say exactly that the real tension is less than or equal to the BPS tension  $T(n) \leq 4\pi |n\T|$. This is because the BPS vortex is not a solution of the equations of motions and so it is not the one that minimizes the tension (\ref{nonBPS}). Note also that the first two corrections (\ref{onecor}) and (\ref{twocor}) are positive, which  means that the last one (\ref{threecor}) is negative and must dominate.

To decide whether the vortex is of type  \textrm{I} or  type \textrm{II}, one should verify how the first order  correction scales with $n$.  Looking at the equations for the profile functions (\ref{profileeq}), we see that $R_v$ grows with $n$, but 
this information is not enough to establish whether the sum of the three corrections grows or decay, because they scale differently with $R_v$ (compare for example (\ref{twocor}) with (\ref{threecor})). To get this information we must know the specific superpotential (see for example \cite{VYung} and \cite{fuertes}). If the correction grows with $n$ then $T(2)<2 \, T(1)$ and the vortices are of type \textrm{I}, on  the contrary they are of type \textrm{II}.

\section{ Summary and    Discussion \label{sec:conclusion}      }

Here we summarize the main result of this paper. We have studied the tension of vortices that arise in color-flavor locking vacua  of $\N=2$ SQCD broken to $\N=1$ by a superpotential. The tension can be written as
\beq
\label{claimconc}
T=4\pi|\T| + T_{non\, BPS}\ ,
\eeq
where $\T$ is the  holomorphic tension and is related to the central charge of the theory. To compute $\T$ we must first solve the factorization equations:
\beq
\label{factoroneconc}
{y}^2=P_{N_c}(z)^2-\Lambda^{2N_c-N_f}\prod_{I=1}^{N_f}(z-m_I)=
F_{2n}(z)H_{N_c-n}(z)^2 
\eeq
and
\beq
\label{factortwoconc}
{y_{m}}^2=W\p(z)^2+f(z)={g_{k}}^2 F_{2n}(z) Q_{k-n}(z)^2 \ .
\eeq
These equations give, as a function of $W(z)$ and $\Lambda$, the points of the moduli space that survive after the perturbation and the polynomial $f(z)$. In particular we obtain $\tm_I$, that is the double root of $H_{N_c-n}(z)^2$ in (\ref{factoroneconc}) that in the semiclassical limit becomes $m_I$. With these results we can compute the holomorphic tension of the $I$-vortex: 
\beq
\label{strongtensionconc}
\T_I=\left.\sqrt{{W\p}^2+f}\right|_{z=\tm_I}\ .
\eeq

What is the nature of the quantum corrections to the holomorphic tension?
The corrections comes from $\tm$ and $f$ and both of them are computed using the factorization equations (\ref{factoroneconc}) and (\ref{factortwoconc}).  When $\Lambda = 0$ the computation gives $\tm=m$ and $f=0$, thus the corrections are positive powers of  $\Lambda$. Inserting in (\ref{strongtensionconc}) we can expand in powers of $\Lambda /m$:
\beq
\T = W\p(m) + \sum_{l=1}^{\infty} \, \T_l \left(\frac{\Lambda}{m}\right)^l \ .
\eeq
The terms $\T_l$ can be interpreted, in the semiclassical limit, as an $l$-instantons correction to the holomorphic tension and the  perturbative corrections are absent.

In the semiclassical region our vortices are created by the winding of the quarks $Q$,  $\tQ$ and so they carry magnetic flux. In the strong coupling region the vortices are created by the winding of dual quarks and so they carry also electric flux.
As $m$ is varied and
reaches values below the dynamical scale
$\Lambda$  of the theory the change of monodromy around the quark singularity  occurs, when it  moves below the 
cuts produced by other singularities.  Quarks may become magnetic monopoles \cite{SW2}.  The precise way this type of metamorphosis takes place
has been studied explicitly  only in the simplest  cases \cite{CV}.   
For this reason we believe that, in the same way quarks are continuously changed in the dual quarks, the color-locked vortices are continuously connected if we move the parameter $m$. The same problem was encuontered in \cite{miofirst} dealing with non-abelian magnetic monopoles.

There is also a non-BPS contribution to the tension. We have found that a region of parameters exists where the non-BPS corrections are small with respect to the BPS tension. In this region these vortices are almost BPS.  An interesting question still unsolved is:   can one find a condition to distinguish whether the vortices are of type \textrm{I} or  type \textrm{II} ?

Our analysis has been performed in the strong coupling regime where there are two different scales of symmetry breaking
\beq
U(N_c) \longrightarrow U(1)^{N_c} \longrightarrow U(1)^{n} \ .
\eeq
The topological stability of the $N_c-n$ vortices is assured only at the lower energy scale by the homotopy group  $\pi_1(U(1)^{N_c-n})=\mathbb{Z}^{N_c-n}$. 
If we relax the strong coupling condition  a vortex can decay in a lighter one. Out of the strong coupling regime our computation breaks down for two reasons. First the dual-quark condensate has no meaning out of the strong coupling regime and thus we don't know how to compute the holomorphic tension.\footnote{In \cite{Stefano2}, dealing with nonabelian vortices, we have found an expression for the holomorphic tension valid out of the strong coupling, but here unfortunately we have no more control of the non-BPS corrections.} Second the non-BPS corrections become in general large and are no more under control.

Moreover, there is a another issue we want to discuss.
The color-flavor locking points are not the only ones where there are massless charged particles that condense and create a vortex. Consider, for example, pure $U(N_c)$ theory \cite{SW1,DS}, in which there are points where some dyons become massless and condense. What about the tension in this case? The analysis at strong coupling given in section \ref{sec:check} is still valid. In this case one cannot bring these kind of vacua at weak coupling because the dyons are always massive in this regime.

Finally we consider the simplest example $n=k=0$, where the superpotential is linear 
\beq
W(z)=g_0 z \ .
\eeq 
As ${W\p}\p=0$, there is no lost of information in rotating away the superpotential so that it becomes an  FI term for the global $U(1)$:  $-\int d^2\theta
d^2 \bar{\theta} 2r \Tr_{N_c}\, V$. This is the same model considered in \cite{tong-monopolo} and \cite{hanany-tong}.\footnote{ Similar vortices are found also in supersymmetric theories in six spacetime dimensions
with fundamental hypermultiplets \cite{nitta}.} Classically the vacua that survives are the ones completely locked, so we need at least $N_c$ flavors. The holomorphic tension is 
\beq
\label{simpleconc}
\T=g_0
\eeq
and the non-BPS contribution is absent. Our result is particularly powerful in this case, in fact the polynomial $f(z)$, being of degree $z-1$, is zero. So there are no quantum corrections to (\ref{simpleconc}) and also the non-BPS contribution is absent for every value of $m$. Thus the vortices are exactly BPS and the tension is exactly $T=4\pi|g_0|$.

\appendix

\section{Details on the central charge }   \label{centralcharge}

The holomorphic tension is strictly related to the central charge of the theory, as in the case of the dyon  mass  in $\N=2$.  Here we review the tools to study the central charge in the presence of a vortex \cite{GS}. In the presence of a vortex the $\N=1$ superalgebra is modified by  of a term that breaks the Lorentz invariance
\bea
&\left\{Q_{\alpha},{\overline{Q}}_{\dot{\a}}\right\}=2P_{\alpha\dot{\a}}+2Z_{\alpha\dot{\a}\ ,}& \\
& \big\{Q_{\alpha},Q_{\a}\big\}=0\ ,\qquad \big\{{\overline{Q}}_{\dot{\alpha}},{\overline{Q}}_{\dot{\a}}\big\}=0 \ .& \nonumber
\eea
The central charge $Z_{\mu}$ is proportional to the vortex orientation \footnote{$n_{\mu}$ is the vortex spatial orientation in the rest frame.} $n_{\mu}$  multiplied by its length $L$
\beq
\label{tensionprop}
Z_{\mu}=\t L n_{\mu}\ ,
\eeq
where $\t$ is a constant of proportionality. 
Now we go in the rest frame of this object (the piece of vortex of length $L$), where $P^{\mu}=(M,0,0,0)$ and,  orienting  it in the $\hat{z}$ direction,  $n^{\mu}=(0,0,0,1)$. The algebra becomes
\beq
\left\{Q_{\alpha},{\overline{Q}}_{\dot{\a}}\right\}=2\left(
\begin{array}{cc}M+ \t L&\\&M- \t L\end{array}\right)_{\alpha \dot{\a}}\ .
\eeq
When the mass saturates the bound
\beq
M=T L\ ,
\eeq
half of the supersymmetries are unbroken, and so we see that $T$ is the BPS tension.

The central charge depends on the theory and to calculate it one uses the supercurrent ${\overline{S}}_{\mu\dot{\alpha}}$ that corresponds to the generator
\beq
{\overline{Q}}_{\dot{\alpha}}=\int d^3x \,{\overline{S}}_{0\dot{\alpha}} \ .
\eeq
We calculate the supersymmetric variation of this current
\beq
\label{variation}
\delta_{\alpha}{\overline{S}}_{\nu\dot{\a}}=2\sigma^{\rho}_{\phantom{\rho}\alpha\dot{\a}}T_{\nu\rho}+\partial^{\rho}R_{\nu\rho\,\alpha\dot{\a}} \ ,
\eeq
where $T_{\nu\rho}$ is the energy-momentum tensor and $\partial^{\rho}R_{\nu\rho\,\alpha\dot{\a}}$ is a boundary term that satisfy
\beq
R_{\nu \rho \,\alpha \dot{\a}}=-R_{\rho \nu \,\alpha\dot{\a}} \ .
\eeq
Integrating (\ref{variation}) we obtain the superalgebra
\beq
\int d^3x\,\delta_{\alpha}{\overline{S}}_{0\dot{\a}}=
\Big\{Q_{\alpha},\int d^3x\,{\overline{S}}_{0\dot{\a}}\Big\}=
2P_{\alpha\dot{\a}}+2Z_{\alpha\dot{\a}} \ .
\eeq
Thus, the central charge is given by the integral of the boundary term in (\ref{variation})
\beq
\label{integralZ}
2Z_{\alpha\dot{\a}}=\int d^3x \,  \partial^iR_{0i\,\alpha\dot{\a}} \ .
\eeq
To compute $R$, it is convenient to go in bispinorial  notation.\footnote{ If $v_{\mu}$ is a vector, the passage is given by these formulas: $v_{\a\dot{\a}}=\sigma^{\mu}_{\phantom{\mu}\a\dot{\a}}v_{\mu}$ and $v^{\mu}=-\frac{1}{2}\bar{\sigma}^{\mu\a\dot{\a}}v_{\a\dot{\a}}$.}  The variation (\ref{variation})  in bispinorial notation is
\beq
\label{variationbispi}
\delta_{\alpha}{\overline{S}}_{\beta\dot{\beta}\dot{\a}}=2 T_{\beta\dot{\beta}\,\a\dot{\a}}+\partial^{\rho}R_{\beta\dot{\beta}\,\rho\,\alpha\dot{\a}} \ .
\eeq
The two terms can be distinguished by symmetry: the energy momentum tensor contains the terms both symmetric in $\a\,\beta$ and $\dot{\a}\,\dot{\beta}$ or both antisymmetric, while $R$ contains the terms with  mixed symmetry.   Now we consider an $\N=1$ SQED with gauge multiplet  $A_{\mu},\lambda$ and some charged chiral  multiplets $q_i,\psi_1$ of charge $t_i$. The computation of $R$ in this model gives 
\beq
R_{\beta\dot{\beta}\phantom{\rho}\alpha\dot{\a}}^{\phantom{\beta\dot{\beta}}\rho} \propto \sum_i \,
(\epsilon_{\dot{\a}\dot{\beta}}\sigma^{\rho\nu}_{\phantom{\rho\nu}\a\beta}+ \epsilon_{\a\beta}\bar{\sigma}^{\rho\nu}_{\phantom{\rho\nu}\dot{\a}\dot{\beta}})\, A_{\nu}q\dag_it_iq_i  + ferm \ .
\eeq
We don't need to consider the fermion terms because in the vortex solution  only the bosonic fields are excited.
Coming back to vector notation, we get
\beq
R_{\nu\rho\mu} =  \sum_i \, \epsilon_{\nu\rho\mu\tau} A^{\tau} q\dag_it_iq_i 
\eeq
and (\ref{integralZ}) becomes:
\beq
Z_0=0 \ , \qquad Z_j \propto \int d^3x \, \sum_i \, \de^l (\epsilon_{ljk}A_kq\dag_it_iq_i)\ .\eeq
Remembering (\ref{tensionprop}) one obtains the tension of the vortex
\beq
\label{ultimino}
T_{BPS} =  \oint d\vec{x} \cdot \vec{A} \, \sum_i \, q\dag_it_iq_i \ .
\eeq

The term that appears in the BPS tension is the same that appears in the $D$ term  of  the potential 
\beq
\label{ultiminoino}
V_D=\frac{e^2}{2}(\sum_i \, q\dag_it_iq_i-2r)^2 \ ,
\eeq
where we have considered the option of a Fayet-Iliopulos term in the Lagrangian:\beq
-\int d^2\theta
d^2 \bar{\theta} \, 2r V \ .
\eeq
The FI  term is crucial because without it the central charge would be zero, in fact from (\ref{ultimino}) and (\ref{ultiminoino})  the BPS tension  is proportional to the FI term:
\beq
\label{BPScentraltension}
T_{BPS} =  2r \oint d\vec{x} \cdot \vec{A} \ .
\eeq

\section{Note on conventions} \label{app:conventions}

In the conventions that we use in the paper (\ref{QFT})
\beq
\left.W^{\a}W_{\a}\right|_{\theta\theta}=-\frac{1}{2}F_{\mu\nu}F^{\mu\nu}-2i\bar{\lambda}\bar{\sigma}^{\mu}D_{\mu}\lambda+D^2+\frac{i}{4} \epsilon_{\mu\nu\rho\tau}F_{\mu\nu}F_{\rho\tau}\ .
\eeq
To have the same normalization of \cite{CSW} we should use the following definitions:                
\beq
T(z)=\Tr\, \frac{1}{z-\Phi}\ ,\qquad
R(z)=-\frac{1}{16\pi^2}\Tr\,\frac{W^{\a}W_{\a}}{z-\Phi} \ ,
\eeq
\beq
M_I(z)=\tQ_I \frac{1}{z-\Phi}Q_I \ .
\eeq
Taking into account that our superpotential is $\sqrt{2}({\widetilde Q}\Phi Q-m{\widetilde Q}Q+W(\Phi))$, the generalized Konishi anomalies of \cite{CSW} become:
\bea
\label{CDSWKenanomaly}
&[\sqrt{2}W\p(z)R(z)]_-= R(z)^2 \ ,\\
&[M_I(z) \sqrt{2} (z-m_I)]_-= R(z) \ .\nonumber
\eea 
In the article we use, instead of this, the convection  (\ref{convenction}) for $R(z)$, such that  (\ref{CDSWKenanomaly}) are simplified and become (\ref{kenanomaly}).

\section* {Acknowledgements}

 I would like to  thank  people from whom I learned a lot of physics and mathematics used in this paper: Jarah Evslin, Kenichi Konishi, Marco Matone and Alexei Yung. I thank also  for useful discussions:  Roberto Auzzi, Sergio Benvenuti, Giacomo Marmorini and  Erik Tonni.
Finally I thank the organizers and the lectures of the ``Introductory school on recent developments in supersymmetric gauge theories 2004'' at ICTP Trieste.

\end{document}